\documentstyle[12pt,fa4]{article}
\newcommand{\be}{\begin{equation}}
\newcommand{\ee}{\end{equation}}
\newcommand{\bea}{\begin{eqnarray}}
\newcommand{\ena}{\end{eqnarray}}
\newcommand{\vs}[1]{\rule[- #1 mm]{0mm}{#1 mm}}
\newcommand{\hs}[1]{\hspace{#1mm}}
\newcommand{\mb}[1]{\hs{5}\mbox{#1}\hs{5}}

\newcommand{\sm}[2]{\frac{\mbox{\footnotesize #1}\vs{-2}}
                   {\vs{-2}\mbox{\footnotesize #2}}}
\newcommand{\prt}{\partial}

\newcommand{\Z}{Z\hspace{-2mm}Z}

\newcommand{\po}{{\mbox{\small{$\ast$}}}}

\newcommand{\half}{\frac{1}{2}}
\newcommand{\wt}[1]{\widetilde{#1}}
\newcommand{\sect}[1]{\setcounter{equation}{0}\section{#1}}

\newcommand{\cg}{{\cal G}}
\newcommand{\cd}{{\cal D}}

\newcommand{\r}[1]{{\cal R}_{#1}}
\newcommand{\rpi}[1]{{\cal R}_{#1}^{\pi}}
\newcommand{\tg}{{\widetilde{\cal G}}}

\newcommand{\PL}[1]{Phys.\ Lett.\ {\bf #1}}

\newcommand{\CMP}[1]{Comm.\ Math.\ Phys.\ {\bf #1}}

\newcommand{\MPL}[1]{Mod.\ Phys.\ Lett.\ {\bf #1}}

\begin{document}
\renewcommand{\thefootnote}{\fnsymbol{footnote}}

\newpage
\setcounter{page}{0}
\pagestyle{empty}

\vs{30}

\begin{center}

{\LARGE {\bf (Super) $W$-algebras from non Abelian}}\\[.5cm]
{\LARGE {\bf (super) Toda theories}}\\[1cm]

\vs{10}

{\large
L. Frappat\footnote{Groupe
d'Annecy, LAPP Chemin de Bellevue BP 110, F-74941 Annecy-le-Vieux Cedex,
France}, E. Ragoucy${}^{\po}$
and P. Sorba${}^{\po}$\footnote{Groupe de Lyon, ENS Lyon 46 all\'ee
d'Italie, F-69364 Lyon Cedex 07, France}}\\[.5cm]
{\em Laboratoire de Physique Th\'eorique }
{\small E}N{\large S}{\Large L}{\large A}P{\small P}
\footnote{URA 14-36 du
CNRS, associ\'ee \`a l'E.N.S. de Lyon, et au L.A.P.P. (IN2P3-CNRS)}\\
\end{center}
\vs{20}

\centerline{ {\bf Abstract}}

\indent

A classification of (super) $W$ algebras arising from non Abelian Toda and
super Toda theories is presented. This classification is based on the
$Sl(2)$ or $OSp(1|2)$ sub(super)algebras of the simple Lie
(super)algebra underlying the model. This allows to compute the conformal
spin content of each $W$ (super)algebra.

\indent

{\em Based on two lectures given by L. Frappat and E. Ragoucy at the\\
XIX International Colloquium on Group Theoretical
Methods in Physics\\
Salamanca (Spain), June 29 - July 4, 1992}

\newpage
\pagestyle{plain}

\sect{Bosonic case}

\noindent

{\large{\bf From WZW to Abelian Toda models}}

\indent

As a pedagogical introduction, we present the basic ideas that are
used to go from a WZW model to the usual (so-called Abelian) Toda
theory \cite{Sav,ORaf}.
We start with a WZW action defined on a connected real Lie group
$G$, of Lie algebra $\cg$:
\bea
S(g) &=& \frac{\kappa}{2}\int d^2x <g^{-1}\prt_+g,g^{-1}\prt_-g>
\nonumber\\
&& +\frac{\kappa}{2}\int d^2x\int dt
<g^{-1}\prt_tg,(g^{-1}\prt_+g)(g^{-1}\prt_-g)
+(g^{-1}\prt_-g)(g^{-1}\prt_+g)> \label{aWZW}
\ena
The equations of motion
$\prt_-(g^{-1}\prt_+g)=0$ and $\prt_+(g^{-1}\prt_-g)=0$
are just the conservation laws of the Kac-Moody (KM)
currents $J_\pm=g^{-1}\prt_\pm g$. We will focus hereafter on the $J_-$.

\indent

The basic idea is to impose constraints on some components of
these currents, namely:
\be
J_-|_{\alpha<0}\equiv M_-=\sum_{i=1}^r \mu_iE_{-\alpha_i} \ \
\forall\alpha\in\Lambda \mb{with} \mu_i>0\label{cont1}
\ee
where $\Lambda$ is the set of all roots and $\{\alpha_i,\
i=1,..,r=\mbox{rank}\cg\}$ a simple roots system of $\cg$.
Such constraints can be obtained as a part of the equations
of motion of a new model resulting from a Lagrange multiplier treatment
on the action (\ref{aWZW}).
Then, a gauge theoretical approach, and the use of the (local)
Gauss decomposition $g=g_+hg_-$, with
\be
g_\pm=\exp(\sum_{\alpha>0}\Gamma^{\pm\alpha}E_{\pm\alpha}) \mb{and}
h=\exp(\phi^iH_i)
\ee
provide in the Euler equations, the Toda differential equations
in the $\phi_i$ variables.

\indent

The stress
energy tensor $T$ of this theory can be easily deduced from the
original WZW one $T_0$ by looking at the conformal properties of the
currents. In fact, the constraints (\ref{cont1}) break the original
conformal invariance, for we set to constant some components of the
vector field (i.e. of the conformal spin 1 primary field) $J$. To
restore this conformal invariance, we have to modify the conformal
spin of the components which are set to constant. For such a purpose,
one introduces the gradation $H$ defined as
${[H,E_{\alpha_i}]}= E_{\alpha_i}$, $i=1,..,r$.
Then, from the eigenspaces decomposition of $\cg$ with respect to $H$:
\be
\cg=\oplus_h\cg^{(h)} \mb{with} [H,X^h]=h X^h\ ,\ \forall
X^h\in\cg^{(h)}
\ee
it is clear that $T=T_0+<H,\prt J>$,
so that for $J^{h}=<X^{h},J>$ we have
\[
\{T(z),J^{h}(w)\}_{PB} = (1+h)\, J^{h}(w)\prt\delta(z-w)
+(\prt J^{h})(w)\delta(z-w) + <H,X^{h}>\prt^2\delta(z-w)
\]
which proves that $J^{h}$ has spin $1+h$, and is primary as soon as
$<H,X^{h}>=0$.

\indent

The residual gauge transformations
$J\rightarrow J^g=g_+ J g_+^{-1}+(\prt
g_+)g_+^{-1}$
allow to transform $J$ into a gauge invariant current
\be
J^g=M_-+\sum_{j\geq0}W_{j+1}(J)M_j \mb{with} [M_-,M_j]=0
\ee
where $W_{j+1}$ are polynomials in the
currents. These fields are "carried" by the generators $M_j$, all
orthogonal to $H$, so that the $W_{j+1}$'s are primary and
of conformal spin $j+1$.
These quantities close under the Poisson Brackets induced by
the KM currents, and form the $W$ algebra associated to $\cg$.
The expressions of the $W_{j+1}$'s as functions of the original
currents $J$ give an explicit realization of the $W$ algebra.
The conformal spin content of this $W$ algebra is directly
given by the adjoint decomposition of $\cg$ with respect to the
$Sl(2)$ subalgebra formed by $\{M_+,H,M_-\}$, where $M_+$ is the
unique (up to a conjugation) element such that $\{M_+,H,M_-\}$ is an
$Sl(2)$ algebra. Note that the $M_j$'s are the lowest weights of
this $Sl(2)$ algebra. In the case of Abelian Toda (presented in this
section), this subalgebra is nothing but the principal $Sl(2)$ algebra
of $\cg$, so that the adjoint decomposition makes appear the exponents
of $\cg$: we recover that the spins of the (usual) $W$ algebra
associated to $\cg$ are just the exponents plus one.

\indent

{\large{\bf Generalization}}

\indent

They are essentially two ways of generalization:

$\bullet$ Choose different components to be set to constant.

$\bullet$ Make a supersymmetric treatment of the method.

\indent

We will present here the first generalization, the second
one being presented by L. Frappat in the following contribution.

\indent

As it
should be clear to the reader, as far as one wants to preserve a
conformal invariance, changing the components which are set to
constant is the same thing as choosing
another gradation in $\cg$. This gradation must satisfy a {\em non
degeneracy} condition which ensures that the $W$ fields will be of
positive conformal spin. For any decomposition
\be
\cg=\oplus_h \cg^{(h)} \mb{and} J=\sum_h J^{h}
\ee
the constraints will be
\be
J^{(-1)}= M_- \mb{and} J^{h}=0 \mb{if} h<-1
\ee
In that way, one obtains {\em non Abelian Toda Theory} and new $W$
algebras, which are classified by the different $Sl(2)$ subalgebras of
$\cg$.

We start with a gradation $H$ of $\cg$, and impose the constraints
(\ref{cont1}). Then, the stress energy tensor is changed to
$T=T_0+<H,\prt J>$, and the residual gauge transformations allow to
transform $J$ into $J^g=M_-+\sum_j W_{j+1}M_j$, where $W_{j+1}$ are
gauge invariant polynomials. To get a $W$ algebra, one has to check
that these fields are primary. This is
true only when the gradation $H$ is the Cartan generator $M_0$ of the
$Sl(2)$ algebra generated by $M_-$, as it was in the Abelian
case. If $H$ differs from $M_0$, the $W$ algebra is in fact isomorphic
to the one obtained by a gradation with $M_0$, but with a "wrong"
stress energy tensor \cite{OR2}. This isomorphism proves that
the $W$ algebras are classified by the $Sl(2)$ embeddings,
but also that the $U(1)$ factor $Y=H-M_0$ provides a conserved
hypercharge which is specially helpful
(by its conservation) to calculate the Poisson
Brackets of the algebra \cite{bouc}. Of course, due to the non degeneracy
condition, the $U(1)$ factor is of special type: its eigenvalues must
obey to the rule $|y|\leq j$ for any $Sl(2)\oplus U(1)$ representation
$D_j(y)$ entering in the decomposition of the adjoint of $\cg$ \cite{pl}.

\indent

{\large{\bf Classification of non Abelian Toda theories}}

\indent

{}From the arguments previously developped,
the different Toda theories are determined by the following method:

$i)$ Classify all the $Sl(2)$ subalgebras of $\cg$

$ii)$ Add eventually an extra $U(1)$ factor with the non degeneracy
condition.

$iii)$ The $Sl(2)\oplus U(1)$ decomposition of the adjoint
representation of $\cg$ gives the spin content and the hypercharge of
the corresponding $W$ algebra.

\indent

The different $Sl(2)$ subalgebras of a simple Lie algebra have been
classified by Dynkin \cite{Dyn}. Up to (known) exceptions
occuring in $D_n$ and
$E_{6,7,8}$, they are all principal $Sl(2)$ embeddings of regular
subalgebras of $\cg$.
For each subalgebras $\wt{\cg}$ of an algebra $\cg$,
one can easily compute the
decomposition of the fundamental representation:

\indent

$\begin{array}{ccc}
\cg & \wt{\cg} & \mbox{Fund. Rep. of \ } \cg \\ \hline
Sl(n)
 &Sl(p) &\cd_{(p-1)/2} \oplus (n-p)\cd_{0} \\
 \hline
Sp(2n)
 &Sp(2p)
 & \cd_{p-1/2} \oplus (2n-2p)\cd_{0} \\
 &Sl(p)_2
 & 2\cd_{(p-1)/2} \oplus 2(n-p)\cd_{0} \\
 &Sl(2)_1
 & \cd_{1/2}\oplus (2n-2)\cd_0 \\
 \hline
SO(2n)
 & SO(2p+2)
 &\cd_{p} \oplus (2n-2p-1)\cd_{0} \\
 &Sl(p)
 & 2\cd_{(p-1)/2}\oplus 2(n-p)\cd_{0} \\
 &Sl(2)_2
 & \cd_1\oplus (2n-3)\cd_{0} \\
 &Sl(4)
 &2\cd_{3/2} \oplus (2n-8)\cd_{0} \\
 &SO(6)
 & \cd_2\oplus (2n-5)\cd_{0} \\
 &2Sl(2)
 &4\cd_{1/2} \oplus (2n-8)\cd_{0} \\
 &SO(2k+1)\oplus SO(2n-2k-1)
 & \cd_k\oplus \cd_{n-k-1} \\
\hline
SO(2n+1)
 & SO(2p+1)
 &\cd_{p} \oplus (2n-2p)\cd_{0} \\
 &Sl(p)
 & 2\cd_{(p-1)/2}\oplus (2n-2p+1)\cd_{0} \\
 &Sl(2)_2
 & \cd_1\oplus (2n-2)\cd_{0} \\
 &Sl(4)
 &2\cd_{3/2} \oplus (2n-7)\cd_{0} \\
 &SO(6)
 & \cd_2\oplus (2n-4)\cd_{0} \\
 &2Sl(2)
 &4\cd_{1/2} \oplus (2n-7)\cd_{0} \\
\end{array}$

\indent

Then, the adjoint representation is obtained by the product of the
fundamental representation by its contragredient. For $SO(n)$ and
$Sp(2n)$ algebras, one has to use the formulae:
\bea
\left( \cd_n\times \cd_n\right)_A &=& \cd_{2n-1}\oplus \cd_{2n-3}\oplus \cdots
\oplus \cd_1 \hs{5} n\in \Z \nonumber\\
\left( \cd_{n-(1/2)}\times \cd_{n-(1/2)}\right)_A &=&
\cd_{2n-2}\oplus \cd_{2n-4}\oplus \cdots\oplus \cd_0 \hs{5}
n\in \Z \nonumber\\
\left( \cd_n\times \cd_n\right)_S &=&
\cd_{2n}\oplus \cd_{2n-2}\oplus \cdots\oplus \cd_0 \hs{5}
n\in \Z \nonumber\\
\left( \cd_{n-(1/2)}\times \cd_{n-(1/2)}\right)_S &=&
\cd_{2n-1}\oplus \cd_{2n-3}\oplus \cdots
\oplus \cd_1 \hs{5} n\in \Z \nonumber
\ena
the subscript (A) S standing for (Anti-)Symmetric part of the product. We have
also, for $m,p\in\Z$ and $j,k\in\half\Z$:
\bea
\left\{\rule{.0mm}{4mm} (m\cd_j)\times (m\cd_j)\right\}_A &=&
\sm{$m(m+1)$}{2} \left( \cd_j\times
\cd_j\right)_A \oplus\sm{$m(m-1)$}{2} \left( \cd_j\times
\cd_j\right)_S \nonumber\\
\left\{\rule{.0mm}{4mm} (m\cd_j)\times (m\cd_j)\right\}_S &=&
\sm{$m(m+1)$}{2} \left( \cd_j\times
\cd_j\right)_S \oplus\sm{$m(m-1)$}{2} \left( \cd_j\times
\cd_j\right)_A  \nonumber\\
\left\{\rule{.0mm}{4mm} (m\cd_j)\times(p\cd_k)\oplus
(p\cd_k)\times(m\cd_j)\right\}_A
&=& \left\{\rule{.0mm}{4mm} (m\cd_j)\times(p\cd_k)\oplus (p\cd_k)\times(m\cd_j)
\right\}_S \nonumber\\
&=& mp\left( \cd_j\times \cd_k\right) \nonumber
\ena
where $m\cd_j$ stands for the direct sum of $m$ representations $\cd_j$.

\indent

When the $U(1)$ factor is present, one has to use:
\bea
&&\left\{ \rule{.0mm}{5mm} [n\cd_j(y)\oplus n\cd_j(-y)]\times
[n\cd_j(y)\oplus n\cd_j(-y)] \right\}_P
= n^2 \left( \rule{.0mm}{5mm} \cd_j\times \cd_j\right)(0)\oplus
\nonumber\\
&& \hs{14} \oplus\left(\rule{.0mm}{5mm}
(n\cd_j)\times (n\cd_j)\right)_P(2y) \oplus
\left(\rule{.0mm}{5mm} (n\cd_j)\times (n\cd_j)\right)_P(-2y)\
\mbox{for}\ P=A\ \mbox{or}\ S \nonumber\\
&&\left\{ \rule{.0mm}{5mm} 2[n\cd_j(y)\oplus n\cd_j(-y)]\times
[p\cd_k(y')\oplus p\cd_k(-y')] \right\}_A=
(\cd_n\times \cd_p)(y+y')\oplus
\nonumber \\
&&\hs{14} \oplus(\cd_n\times \cd_p)(-(y+y'))\oplus
 (\cd_n\times \cd_p)(y-y')\oplus (\cd_n\times \cd_p)(-(y-y'))
\ena
Looking at the non degeneracy condition (in the $\cg$ adjoint
representation), lead to the following conditions (necessary but not
sufficient) on the $U(1)$ eigenvalues in the fundamental
representation of $\cg$.
For the $A_n$ series, all the $Sl(2)$
representations $\cd_j$
entering in the fundamental of $\cg$
have the same $U(1)$ eigenvalue, for identical $j$.
For the $SO(n)$ (resp. $Sp(2n)$) algebras, the only
$\cd_j$ representations which have a non zero $U(1)$ eigenvalue
are those appearing in the $\cg$ fundamental representation as
$n(\cd_j(y)\oplus\cd_j(-y)$ with $n=1$ and $j$ integer (resp.
half-integer). If this condition is achieved,
$y=\pm\half$ or $\pm\frac{1}{4}$ when
$j\neq0$ (resp. in all cases).

\indent

{\large{\bf Example: case of $SO(8)$}}

\indent

We give in table \ref{t5} the spin content and the hypercharges of the $W$
fields of all the $W$ algebras obtained from $SO(8)$. A complete
classification for algebras of rank up to 4 can be found in \cite{bouc}.

\begin{table}
\begin{tabular}{|c|l|l|} \hline
Subalg. & $Sl(2)\oplus U(1)$ decompos.
& Spin contents \\
&(fundamental rep.) & (Hypercharge) \\ \hline
&& \\
$A_1$
& $2\cd_{1/2} \oplus 4\cd_0$ & $2,8\po \sm{3}{2},9\po 1$ \\
&& \\
$2A_1$ &$\cd_1\oplus 5\cd_0$
& $6\po 2,10\po 1$ \\
&& \\
$(2A_1)'$ &$4\cd_{1/2}$ & $6\po 2,10\po 1$ \\
&& \\
$3A_1$ & $\cd_1\oplus 2\cd_{1/2}\oplus \cd_0$
& $\sm{5}{2},\sm{5}{2},3\po 2,4\po \sm{3}{2},3\po 1$ \\
&& \\
$\left.\begin{array}{c} A_2 \\ 4A_1 \end{array}\right\}$
& $\cd_{1}(y_1)\oplus \cd_{1}(-y_1)\oplus $
& $3,7\po 2,2\po 1$ \\
& $\oplus \cd_{0}(y_0)\oplus \cd_{0}(-y_0)$
& $(0,\pm y_1\pm y_0,\pm2y_1,3\po0)$ \\
&& \\
$A_3$& $2\cd_{3/2}$
& $4,3\po 3,2,3\po 1$ \\
&&\\
$D_3$& $\cd_2 \oplus 3\cd_0$
& $4,3\po 3,2,3\po 1$ \\
&&\\
$B_2\oplus B_1$& $\cd_2\oplus \cd_1$
& $4,4,3,3\po 2$ \\
&&\\
$D_4$ & $\cd_3\oplus \cd_0$ & $6,4,4,2$ \\
&& \\ \hline
\end{tabular}
\caption{$W$ algebras for $D_4$\label{t5}.}
\end{table}

\sect{Supersymmetric case}

We present here a complete classification of the super $W$ algebras arising
in the super-Toda models \cite{Nous}.
It is now well known that the Toda models can be obtained as constrained
WZW models \cite{ORaf,2}.
Since this derivation has been explained in the contribution of E. Ragoucy
in these proceedings for the bosonic case, and
that the corresponding supersymmetric version is straightforward, we will
focus here only on the algebraic aspect of the problem. After presenting
the main steps of the construction, we briefly explain each of them.

Following the method used in the bosonic case, the different (non Abelian)
super-Toda theories can be classified in that way:

$i$) One determines all the $OSp(1|2)$ subsuperalgebras (SSA) of a basic Lie
superalgebra (BSA) $\cg$.

$ii$) An $OSp(1|2)$ SSA of $\cg$ being given, one decomposes the adjoint
representation of $\cg$ under this $OSp(1|2)$: $\cg = \bigoplus_{i \in I}
\r{j_i}$, where $\r{j_i}$ are $OSp(1|2)$ irreducible representations (IR),
which decompose under the bosonic $Sl(2)$ of $OSp(1|2)$ as the irreducible
representations $\cd_{j_i} \oplus \cd_{j_i+1/2}$ of $Sl(2)$.
For the infinite series $A(m,n)$, $B(m,n)$, $C(n+1)$ and $D(m,n)$, the
$\cg$ decomposition can be obtained from the decomposition of the $\cg$
fundamental representation.

$iii$) One adds to each $OSp(1|2)$ subsuperalgebra a commuting $U(1)$ factor
such that, in the decomposition of $\cg$ into $OSp(1|2) \oplus U(1)$
representations $\r{j_i}(y_i)$, the following non-degeneracy conditions hold:
\be
|y_i| \le j_i \mb{with} i=1,...,n \mb{and} j_i + y_i \in \Z \mbox{ or } \Z+1/2
\ee

$iv$) The superspin content of the super $W$ generators of the super $W$
algebra is given by $s_i = j_i + y_i + 1/2$.

\indent

The answer to the point ($i$) is summarized as follows:

{\em Any $OSp(1|2)$ subsuperalgebra in a basic simple Lie superalgebra $\cg$
can be considered as the superprincipal $OSp(1|2)$ SSA of a regular SSA $\tg$
of $\cg$, up to the following exceptions:

1) For $\cg = OSp(2n \pm 2|2n)$ with $n \ge 2$, besides the superprincipal
$OSp(1|2)$ subsuperalgebras described above, there exist $OSp(1|2)$
subsuperalgebras associated to the singular embeddings
$OSp(2k \pm 1|2k) \oplus OSp(2n-2k \pm 1|2n-2k) \subset OSp(2n \pm 2|2n)$
with $1 \le k \le n-1$.

2) For $\cg = OSp(2n|2n)$ with $n \ge 2$, there exist $OSp(1|2)$
subsuperalgebras associated to the singular embeddings
$OSp(2k \pm 1|2k) \oplus OSp(2n-2k \mp 1|2n-2k) \subset OSp(2n|2n)$
with $1 \le k \le n-1$.}

The BSA's admitting a superprincipal embedding belong to the following list:
$Sl(n|n \pm 1)$, $OSp(2n \pm 1|2n)$, $OSp(2n|2n)$, $OSp(2n+2|2n)$ and
$D(2,1;\alpha)$ with $\alpha \ne 0, \pm 1$ \cite{Serga}.
Note that these BSA's are the only ones which can be used to build Abelian
super Toda theories \cite{DRS}.

\indent

Now we explain how to solve the point ($ii$). Let us consider the decomposition
of the fundamental representation ${\bf F}$ of $\cg$ under some $OSp(1|2)$
subsuperalgebra, superprincipal embedding of $\tg \subset \cg$:
\be
{\bf F} = \left(\rule{.0mm}{4mm} \oplus_i n_i \r{i} \right)
\oplus \left(\rule{.0mm}{4mm} \oplus_j n_j \rpi{j} \right)
\label{equa1}
\ee
The $OSp(1|2)$ irreducible representations $\r{i}$ and $\rpi{i}$ are
defined as follows:
if the superalgebra $\cg$ under consideration belongs to the unitary series
$Sl(m|n)$ (resp. to the orthosymplectic series $OSp(M|2n)$), the $OSp(1|2)$
IR is denoted $\r{i}$ if the representation $\cd_i$ comes from the
decomposition of the fundamental of $Sl(m)$ (resp. $SO(M)$), and $\rpi{i}$ if
the representation $\cd_i$ comes from the decomposition of the fundamental
of $Sl(n)$ (resp. $Sp(2n)$).

Then the decomposition of the adjoint representation of $\cg$ reads, for the
orthosymplectic series,
\be
{\bf Ad} =
\left.\left(\rule{.0mm}{4mm} \oplus_i n_i \r{i} \right) \times
\left(\rule{.0mm}{4mm} \oplus_i n_i \r{i} \right) \right|_A \oplus
\left.\left(\rule{.0mm}{4mm} \oplus_j n_j \rpi{j} \right) \times
\left(\rule{.0mm}{4mm} \oplus_j n_j \rpi{j} \right) \right|_S \oplus
\left(\rule{.0mm}{4mm} \oplus_i n_i \r{i} \right) \times
\left(\rule{.0mm}{4mm} \oplus_j n_j \rpi{j} \right)
\ee
where the indices $A$ and $S$ stand for the (anti)symmetric parts of the
products, and for the unitary series,
\bea
&&{\bf Ad} = \left(\oplus_i n_i \r{i} \oplus_j n_j \rpi{j} \right) \times
\left(\oplus_i n_i \r{i} \oplus_j n_j \rpi{j} \right) - \r{0}
\mb{for} Sl(m|n) \mbox{ with } m \ne n \\
&&{\bf Ad} = \left(\oplus_i n_i \r{i} \oplus_j n_j \rpi{j} \right) \times
\left(\oplus_i n_i \r{i} \oplus_j n_j \rpi{j} \right) - 2\r{0}
\mb{for} PSl(n|n)
\ena

Moreover, the explicit decomposition of the fundamental representation of the
superalgebra $\cg$ with respect to some $OSp(1|2)$ is given in the table below.
The first column gives the superalgebra $\cg$ under consideration, the
second column the SSA $\tg$ containing the $OSp(1|2)$ as superprincipal
embedding and the last column the decomposition of the fundamental
representation of $\cg$ with respect to this $OSp(1|2)$.

\indent

$\begin{array}{ccc}
\cg & \tg & \mbox{Fund. Rep. of \ } \cg \\ \hline && \\
Sl(m|n)
 &Sl(p+1|p) &\r{p/2} \oplus (m-p-1)\r{0} \oplus (n-p)\rpi{0} \\ && \\
 &Sl(p|p+1) &\rpi{p/2} \oplus (m-p)\r{0} \oplus (n-p-1)\rpi{0} \\ && \\
 \hline && \\
OSp(M|2n)
 &\left.\begin{array}{c} OSp(2k|2k) \\ OSp(2k-1|2k)^* \end{array} \right\}
 &\begin{array}{c}\rpi{k-1/2} \oplus (2n-2k)\rpi{0} \\
  \oplus (M-2k+1)\r{0}\end{array} \\ && \\
 &\left.\begin{array}{c} OSp(2k+2|2k) \\ OSp(2k+1|2k)^* \end{array} \right\}
 &\begin{array}{c}\r{k} \oplus(M-2k-1)\r{0} \\
  \oplus (2n-2k)\rpi{0}\end{array} \\ && \\
 &Sl(p+1|p)
 &\begin{array}{c}2\r{p/2} \oplus (M-2p-2)\r{0} \\
  \oplus (2n-2p)\rpi{0}\end{array} \\ && \\
 &Sl(p|p+1)
 &\begin{array}{c}2\rpi{p/2} \oplus (2n-2p-2)\rpi{0} \\
  \oplus (M-2p)\r{0}\end{array} \\ && \\
\end{array}$

$^*$ exists only if $M$ is odd.

$\begin{array}{ccc}
\cg & \tg & \mbox{Fund. Rep. of \ } \cg \\
 \hline && \\
OSp(2|2n)
 &OSp(2|2)
 &\rpi{1/2} \oplus \r{0} \oplus (2n-2)\rpi{0} \\ && \\
 &Sl(1|2)
 &2\rpi{1/2} \oplus (2n-4)\rpi{0} \\ && \\
 \hline && \\
OSp(2n+2|2n)
 &\begin{array}{c}OSp(2k+1|2k) \oplus \\ OSp(2n-2k+1|2n-2k) \end{array}
 &\r{k} \oplus \r{n-k} \\ && \\
 \hline && \\
OSp(2n-2|2n)
 &\begin{array}{c}OSp(2k-1|2k) \oplus \\ OSp(2n-2k-1|2n-2k) \end{array}
 & \rpi{k-1/2} \oplus \rpi{n-k-1/2} \\ && \\
 \hline && \\
OSp(2n|2n)
 &\begin{array}{c}OSp(2k+1|2k) \oplus \\ OSp(2n-2k-1|2n-2k) \end{array}
 & \r{k} \oplus \rpi{n-k-1/2} \\ && \\
 &\begin{array}{c}OSp(2k-1|2k) \oplus \\ OSp(2n-2k+1|2n-2k) \end{array}
 & \r{n-k} \oplus \rpi{k-1/2}
\end{array}$

\vs{5}

When one adds the $U(1)$ commuting factor, the decomposition of the fundamental
representation {\bf F} of $\cg$ under $OSp(1|2) \oplus U(1)$ takes the form
for the unitary superalgebras
\be
{\bf F} = \left(\rule{.0mm}{4mm} \oplus_i n_i \r{i}(y_i) \right) \oplus
\left(\rule{.0mm}{4mm} \oplus_j n_j \rpi{j}(y_j) \right)
\ee
representations labelled by the same index having the same value of $y$.
We have also to impose the supertraceless condition
\be
\sum_i n_i y_i - \sum_j n_j y_j = 0
\ee
This leads to the following decomposition of the adjoint representation of
$\cg$ (for the $Sl(m|n)$ case with $m \ne n$):
\be
{\bf Ad} = \left(\oplus_i n_i \r{i}(y_i) \oplus_j n_j \rpi{j}(y_j) \right)
\times \left(\oplus_i n_i \r{i}(-y_i) \oplus_j n_j \rpi{j}(-y_j) \right)
- \r{0}(0)
\ee
where we use, for explicit calculation
\be
\left(\rule{.0mm}{4mm} n_i \r{i}(y_i)\right) \times
\left(\rule{.0mm}{4mm} n_j \r{j}(y_j)\right) =
n_i n_j \bigoplus_{k=|i-j|}^{i+j} \r{k}(y_i+y_j)
\mb{with $k$ integer and half-integer}
\ee
and the same formula for $\rpi{}$ representations.

Consider now the case of the orthosymplectic superalgebras. The
decomposition of the $OSp(M|2n)$ fundamental representation ${\bf F}$ under
a certain $OSp(1|2)$ being given by (\ref{equa1}), one is led, from
the considerations developed in the bosonic case, to allow non zero values
$y$ of the $U(1)$ factor only for the following combinations:

\noindent
- the representation $\r{i}$ appears twice and only twice ($n_i = 2$), and
$i$ is integer,

\noindent
- the representation $\rpi{i}$ appears twice and only twice ($n_i = 2$), and
$i$ is half-integer.

\noindent
Moreover, $y$ can only take the values 0, 1/4 or 1/2 if $i \ne 0$ (which lead
to the values 0, $\pm 1/2$ or $\pm 1$ for the $U(1)$ factor in
the adjoint representation of $\cg$).

Finally, starting from a decomposition of the fundamental representation of
$OSp(M|2n)$ under $OSp(1|2) \oplus U(1)$ of the form
\be
{\bf F} = \left(\bigoplus_i \r{i}(y_i) \oplus \r{i}(-y_i) \right) \oplus
\left(\bigoplus_j \rpi{j}(y_j) \oplus \rpi{j}(-y_j) \right) \oplus
\left(\bigoplus_{i,n_i \ne 2} n_i \r{i}(0) \right) \oplus
\left(\bigoplus_{j,n_j \ne 2} n_j \rpi{j}(0) \right)
\ee
the decomposition of the adjoint representation of $OSp(M|2n)$ is given by
\bea
{\bf Ad} &=&
\left.\left(\bigoplus_i \r{i}(y_i) \oplus \r{i}(-y_i) \bigoplus_{i,n_i \ne 2}
n_i \r{i}(0) \right) \! \times \!
\left(\bigoplus_i \r{i}(y_i) \oplus \r{i}(-y_i) \bigoplus_{i,n_i \ne 2}
n_i \r{i}(0) \right) \right|_A \nonumber \\
&\oplus & \! \! \left.
\left(\bigoplus_j \rpi{j}(y_j) \oplus \rpi{j}(-y_j) \bigoplus_{j,n_j \ne 2}
n_j \rpi{j}(0) \right) \! \times \!
\left(\bigoplus_j \rpi{j}(y_j) \oplus \rpi{j}(-y_j) \bigoplus_{j,n_j \ne 2}
n_j \rpi{j}(0) \right) \right|_S \nonumber \\
&\oplus & \! \!
\left(\bigoplus_i \r{i}(y_i) \oplus \r{i}(-y_i) \bigoplus_{i,n_i \ne 2}
n_i \r{i}(0)\right) \! \times \!
\left(\bigoplus_j \rpi{j}(y_j) \oplus \rpi{j}(-y_j) \bigoplus_{j,n_j \ne 2}
n_j \rpi{j}(0)\right) \nonumber \\
\ena
where the (anti)symmetric products of $\r{}$ and $\rpi{}$ representations
are given by:
\bea
\left(\rule{.0mm}{4mm} \r{i}(y_i) \oplus \r{i}(-y_i)\right) \times
\left.
\left(\rule{.0mm}{4mm} \r{i}(y_i) \oplus \r{i}(-y_i)\right) \right|_A
\! \! &=& \! \!
\left.\left(\rule{.0mm}{4mm} \r{i} \times \r{i}\right) \right|_A(2y_i)
\oplus
\left.\left(\rule{.0mm}{4mm} \r{i} \times \r{i}\right) \right|_A(-2y_i)
\nonumber \\
&&\oplus \ \ (\r{i} \times \r{i})(0)
\ena
and
\bea
\left(\rule{.0mm}{4mm} \rpi{i}(y_i) \oplus \rpi{i}(-y_i)\right) \times
\left.
\left(\rule{.0mm}{4mm} \rpi{i}(y_i) \oplus \rpi{i}(-y_i)\right) \right|_S
\! \! &=& \! \!
\left.\left(\rule{.0mm}{4mm} \rpi{i} \times \rpi{i}\right) \right|_S(2y_i)
\oplus
\left.\left(\rule{.0mm}{4mm} \rpi{i} \times \rpi{i}\right) \right|_S(-2y_i)
\nonumber \\
&&\oplus\ \  (\rpi{i} \times \rpi{i})(0)
\ena

\indent

Let us conclude by giving the example of the superalgebra $B(2,1) \equiv
OSp(5|2)$. The following table gives respectively the different
subsuperalgebras in $B(2,1)$ which admit a superprincipal embedding, the
corresponding decompositions of the fundamental representation and the
superconformal spin content of the super $W$ algebras.

\indent

$\begin{array}{ccc}
\mb{SSA in $\cg$} &\mb{Decomposition of the} &\mb{Superconformal spin}    \\
                  &\mb{fundamental of $\cg$} &\mb{of the $W$ superfields} \\
                  &                          &\mb{(Hypercharge)}          \\
\hline && \\
\left.\begin{array}{c} D(2,1) \\ B(1,1) \end{array} \right\}
 &\r{1}(0) \oplus \r{0}(y) \oplus \r{0}(-y)
 &\begin{array}{c} 2, \sm{3}{2}, \sm{3}{2}, \sm{3}{2}, \sm{1}{2} \\
 (0,2y,-2y,0,0) \end{array} \\ && \\
\left.\begin{array}{c} C(2) \\ B(0,1) \end{array} \right\}
 &\rpi{1/2}(0) \oplus 4\r{0}(0)
 &\sm{3}{2}, 1, 1, 1, 1, \sm{1}{2}, \sm{1}{2}, \sm{1}{2}, \sm{1}{2},
 \sm{1}{2}, \sm{1}{2} \\ && \\
A(1,0)
 &2\r{1/2}(0) \oplus \r{0}(0)
 &\sm{3}{2}, 1, 1, 1, 1', 1', \sm{1}{2}, \sm{1}{2}, \sm{1}{2}
\end{array}$

\vs{5}

Notice that the $W_{j+1/2}$ superfield corresponding to the representation
$\r{j} = \cd_j \oplus \cd_{j-1/2}$ has two component fields $w_{j+1}$ and
$w_{j+1/2}$ of spins $j+1$ and $j+1/2$ respectively. If the representation
$\cd_j$ comes from the bosonic (resp. fermionic) part, $w_{j+1}$ is commuting
(resp. anticommuting), whereas $w_{j+1/2}$ is anticommuting (resp. commuting).
Therefore, if $j$ is integer, the generators $w_{j+1}$ and $w_{j+1/2}$ have
the "right" statistics, whereas they have the "wrong" statistics if $j$ is
half-integer. The superconformal spins denoted with a prime are used in the
case of $W$ superfields obeying to the "wrong" statistics.

\indent

Let us conclude by saying a few words about the superconformal algebras. Our
classification constitutes a well-adapted framework to the study of such
algebras. In particular, it allows us to determine all "quasi-superconformal
algebras" (i.e. superconformal algebras, but with wrong statistics for the spin
$\sm{3}{2}$ fields), the $\Z_2 \times \Z_2$ superconformal algebras, and also
the "quadratic-superconformal algebras" (i.e. superconformal algebras with $N$
supersymmetry charges which close quadratically on Kac-Moody currents).
Tables can be found in reference \cite{Nous}. Note that a classification of
the superconformal algebras has been recently given in reference \cite{Machin}.
We are in accordance with their results, but we find also two new algebras.
Finally let us emphasize that all our results are for the moment purely
classical.

\end{document}